\documentclass[referee]{raa}
\pdfoutput=1
\usepackage{natbib}
\usepackage{amsmath}
\usepackage{graphicx}
\usepackage{comment}
\usepackage{color}
\usepackage{booktabs}
\usepackage{threeparttable}
\usepackage{ulem}
\bibpunct{(}{)}{;}{a}{}{,}
\definecolor{DarkGreen}{rgb}{0.0,0.45,0.0}  

\newcommand{\fig}[1]{Figure~\ref{#1}}
\newcommand{\tbl}[1]{Table~\ref{#1}}
\newcommand{\sect}[1]{Section~\ref{#1}}

\begin{document}
\title{Damped large amplitude oscillations in a solar prominence and a bundle of coronal loops}
\setcounter{page}{1}
\author{Quanhao Zhang\inst{1}, Yuming Wang\inst{1,2}, Rui Liu\inst{1,3}, Chenglong Shen\inst{1,2}, Min Zhang\inst{1,4,5}, Tingyu Gou\inst{1,4}, Jiajia Liu\inst{1}, Kai Liu\inst{1}, Zhenjun Zhou\inst{1,4}, Shui Wang\inst{1}}
\institute{CAS Key Laboratory of Geospace Environment, Department of Geophysics and Planetary Sciences, University of Science and Technology of China, Hefei 230026, China; {\it zhangqh@mail.ustc.edu.cn, ymwang@ustc.edu.cn}\\
\and Synergetic Innovation Center of Quantum Information \& Quantum Physics, University of Science and Technology of China, Hefei, Anhui 230026, China\\
\and Collaborative Innovation Center of Astronautical Science and Technology, China\\
\and Mengcheng National Geophysical Observatory, School of Earth and Space Sciences, University of Science and Technology of China, Hefei 230026, China\\
\and Department of Mathematics and Physics, AnHui JianZhu University, Heifei, 230601}

\abstract{We investigate the evolutions of two prominences (P1,P2) and two bundles of coronal loops (L1,L2), observed with SDO/AIA near the east solar limb on 2012 September 22. It is found that there were large-amplitude oscillations in P1 and L1, but no detectable motions in P2 and L2. These transverse oscillations were triggered by a large-scale coronal wave, originating from a large flare in a remote active region behind the solar limb. By carefully comparing the locations and heights of these oscillating and non-oscillating structures, we conclude that the propagating height of the wave is between 50 Mm and 130 Mm. The wave energy deposited in the oscillating prominence and coronal loops is at least of the order of $10^{28}$~erg. Furthermore, local magnetic field strength and Alfv\'{e}n speeds are derived from the oscillating periods and damping time scales, which are extracted from the time series of the oscillations. It is demonstrated that oscillations can be used in not only coronal seismology, but also revealing the properties of the wave.}
\keywords{Sun: filaments, prominences---Sun: flares---Sun: oscillations---waves}
\authorrunning{Q.-H. Zhang et al. }            
\titlerunning{Damped large-amplitude oscillations}  
\maketitle

\section{Introduction}
\par
Prominence (Filament) oscillations have been observed for a long time \citep[e.g.][]{Kleczek1969a}. They are classified into two groups based on their velocity amplitudes: large-amplitude oscillations with velocity amplitude $\geq$ 20 km~s$^{-1}$  \citep{Tripathi2009a} and small-amplitude oscillations with 2-3 km~s$^{-1}$ \citep{Oliver2002a,Arregui2012a}. In earlier observations, large-amplitude oscillations were caused by Moreton waves \citep{Gilbert2008a,Liu2013a}, coronal waves \citep{Okamoto2004a,Hershaw2011a,Liu2012c}, and nearby flares or jets \citep{Vrvsnak2007a,Li2012a}. The recent observation by \cite{Zhang2014a} revealed that a prominence was triggered to oscillate in large amplitudes by the rising chromospheric fibrils underneath. This procedure, named as "flux feeding", is also a possible trigger of large-amplitude oscillations. So far, there are few observations about large-amplitude oscillations in prominences triggered by waves. With the method of prominence seismology, local physical parameters, such as magnetic field strength, can be extracted from the properties of the oscillations \citep{Isobe2006a,Vrvsnak2007a,Oliver2009a}. By analysing the oscillation of the prominence during its slow rise phase, \cite{Isobe2007a} concluded that prominence seismology based on large-amplitude oscillation is also a diagnostic tool for stability and eruption mechanism of the prominence.
\par
There are also oscillations in coronal loops. Damped oscillations of coronal loops are first discovered by the EUV telescope on board the Transition Region and Coronal Explorer (TRACE) spacecraft \citep{Aschwanden1999a,Nakariakov1999a}, and then further discussed by \cite{Schrijver2002a} and \cite{Aschwanden2002a}. \cite{Nakariakov1999a} concluded that all parts of the loop oscillated transversely and in phase, indicating a kink global standing mode of the loop. There are several damping mechanism for kink oscillations of coronal loops, such as footpoint or side energy leakage \citep{Schrijver2000a}, phase mixing \citep{Heyvaerts1983a,Roberts2000a} and resonant absorption \citep{Ruderman2002a,Ruderman2005a}. It is still an open question as to which mechanisms are working in the damping process. Physical parameters of the oscillations, e.g. periods and damping times, can be used to obtain indirect information on the conditions of the plasma and magnetic field in coronal loops \citep{Nakariakov2001a,Goossens2002a,Arregui2007a}.
\par
Large-scale coronal waves were first observed by the Extreme ultraviolet Imaging Telescope \citep[EIT;][]{Delaboudini`ere1995a} on board the Solar and Heliospheric Observatory \citep[SOHO;][]{Thompson1999a}, and hence are also called "EIT waves". In some papers, coronal waves are introduced as "EUV waves" as well, corresponding to the same phenomena. Coronal waves are commonly interpreted as fast magnetosonic waves \citep{Wang2000a,Ofman2002b}, which are always flare-associated, usually propagating from the flare site isotropically at a typical speed of 200$-$500 km~s$^{-1}$ \citep{Nakariakov2005a}. The properties of the wave vary during the propagation because of the interaction with the coronal magnetic structures \citep{Ofman2002b,Gopalswamy2009a,Veronig2010a}. The studies of coronal waves have shed light on fundamental physical problems in solar physics, such as acceleration of the fast solar wind \citep{Cranmer2007a} and the mechanism of coronal heating \citep{Heyvaerts1983a}. There are close relationships between coronal waves and oscillations of coronal structures. The coronal wave is a possible trigger of the oscillations \citep{Hershaw2011a,Kumar2013a} on one hand, and on the other hand the oscillating parameters reveal the the physical properties of both the wave and the oscillating structures \citep{Gilbert2008a}.
\par
In this paper, we study the oscillations of a prominence and a bundle of coronal loops associated with the coronal wave generated by a large flare. The different parameters of the oscillating and non-oscillating structures reveal the propagating properties of the coronal wave. In the following sections, we establish the locations of the relevant prominences and coronal loops through multi-spacecraft observations(\sect{sec:3d}) to investigate the interaction between the wave and the magnetic structures (\sect{sec:overview}), calculate the oscillating properties of the oscillating structures (\sect{sec:parameters}), and estimate the local magnetic field strength and Alfv\'{e}n speed from the oscillating properties (\sect{sec:esti-mag}). By comparing the locations of the structures, we roughly estimate the propagating height of the wave (\sect{sec:height}). Finally, discussion and conclusion are given in \sect{sec:conclusion}.

\section{Instrument and data}
\label{sec:iad}
The prominences and the coronal loops were observed off the east limb in EUV by the Atmospheric Imaging Assembly \citep[AIA;][]{Lemen2012a} onboard the Solar Dynamics Observatory \citep[SDO;][]{Pesnell2012a}. Images taken by the Extreme Ultraviolet Imager \citep[EUVI;][]{Wuelser2004a}) of the Sun Earth Connection Coronal and Heliospheric Investigation \citep[SECCHI;][]{Howard2008a} imaging package on board the Solar TErrestrial RElations Observatory \citep[STEREO;][]{Kaiser2008a} are utilized to exhibit the propagation of the coronal wave. The prominences appeared as dark filaments in the field of view (FOV) of STEREO's `Behind' spacecraft (STB). Two prominences and two bundles of magnetic coronal loops are analyzed in this paper.
\par

\section{Observations}
\label{sec:observation}

\subsection{3D reconstruction of the structures}
\label{sec:3d}

\begin{figure}
\centerline{\includegraphics[width=0.9\textwidth]{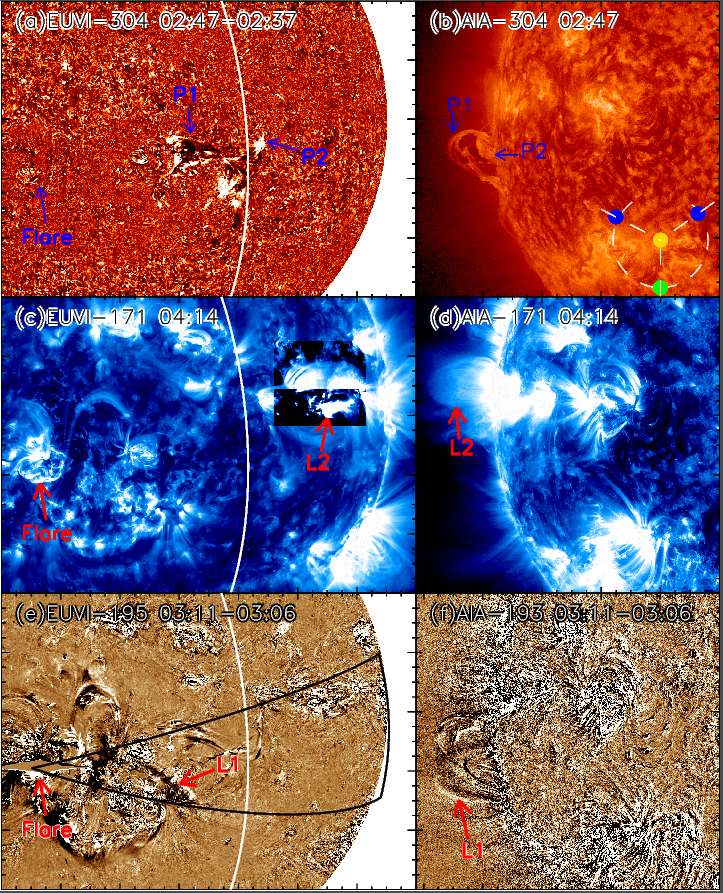}}
\caption{The locations of the prominences and the coronal loops analyzed in this paper. The left panels are images taken by STEREO-B/EUVI, the right panel by SDO/AIA. The white curves in the left panels denote the solar limb as seen by SDO. Panel (a) is the running ratio image of EUVI 304~{\AA} observations; panel (b) is the AIA 304~{\AA} original image, and the inset plots the positions of the STEREO spacecrafts (blue dots) relative to the Sun (yellow dot) and the Earth (green dot) in the plane of Earth's orbit, with STEREO-A ahead of, and STEREO-B behind, the Earth. Panel (c) and (d) are the scaled 171~{\AA} original images from EUVI and AIA, respectively. In panel (c), in order to clearly show the profile of L2, different scalings are used in different regions. Panel (e) is the running difference image of EUVI 195~{\AA} observations, and panel (f) is that of AIA 193~{\AA}. The prominences and the loops are marked as P1, P2, L1 and L2. The solid black lines in the bottom left panel denote the sector region cover both P1 and L1, whose center is located at the flare site.}
\label{fig:location}
\end{figure}

\begin{figure}
\centerline{\includegraphics[width=0.9\textwidth]{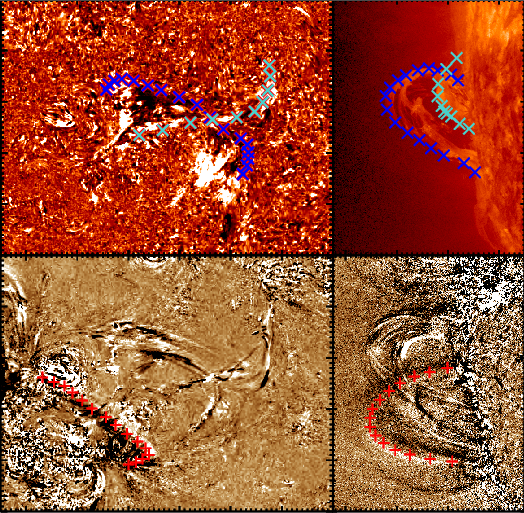}}
\caption{3D reconstructed results of P1, P2 and L1. The images are the same as those in \fig{fig:location}(a), \ref{fig:location}(b), \ref{fig:location}(d), and \ref{fig:location}(f), but zoom in on the region of P1, L1 and L2. The 3D reconstructed points of P1 are marked by blue cross symbols, P2 by cyan cross symbols, and L1 by red pluses.}
\label{fig:3drec}
\end{figure}

\fig{fig:location} illustrates the locations of the relevant prominences and loops from the perspectives of both the SDO and STEREO satellites, the positions of which are shown by the inset in \fig{fig:location}(b). \fig{fig:location}(a) is the running ratio image of EUVI 304~{\AA} observations, and \fig{fig:location}(b) is the original AIA 304~{\AA} images. There are two prominences located at almost the same region in \fig{fig:location}(a), apparently intersecting with each other (also see the animation accompanying \fig{fig:location}), but in fact they are suspended at different heights, as is shown in \fig{fig:location}(b). The higher prominence, spreading along the direction from northeast to southwest in \fig{fig:location}(a), is labelled as `P1', and the lower prominence, generally in east-to-west direction but turns northward at its west end, labelled as `P2'. P1 and P2 can be clearly recognized in the animation accompanying \fig{fig:location}. \fig{fig:location}(c) and \ref{fig:location}(d) are the linearly scaled 171~{\AA} original images from EUVI and AIA, respectively. There is a bundle of coronal loops appearing as a bright dome, which is labelled as `L2'. Different scalings are used in different regions in \fig{fig:location}(c), so as to clearly reveal different parts of L2. \fig{fig:location}(e) is the running difference image of EUVI 195~{\AA} observations, and \fig{fig:location}(f) is the running difference AIA 193~{\AA} image. There is another bundle of coronal loops observed here, called `L1' hereafter. It appears as a dark arch in \fig{fig:location}(e) and \ref{fig:location}(f), as pointed out by the red arrow. L1 is nearer to the flare site than L2. Different from L2, only after the onset of the flare and the passage of a coronal wave would L1 be visible in the FOV of STB (see the animation accompanying \fig{fig:location}), resulting from the oscillating motions triggered by the wave (see \sect{sec:parameters}). 
\par

\begin{table}[htbp]
\centering
\begin{threeparttable}
\caption{Geometrical parameters\label{tbl:gp}}
\begin{tabular}{ccccc}
\toprule
   & {$D$(Mm)*}          & {$H$(Mm)} & {$L$(Mm)**} & {Oscillate or not}\\\hline
P1 & $490^{+80}_{-120}$  & 130       & 220       & Y  \\
P2 & $550^{+80}_{-150}$  & 50        & 80        & N  \\
L1 & $360^{+70}_{-100}$  & 210       & 220       & Y  \\
L2 & $930^{+140}_{-190}$ & 200       & 320       & N  \\\hline
\bottomrule
\end{tabular}
\small 
\begin{tablenotes}
\item[*] The superscripts and subscripts correspond to the spatial range of the structures.
\item[**] $L$ is the estimated length of the parts that can be 3D reconstructed.
\end{tablenotes}
\end{threeparttable}
\end{table}

Observations from different perspectives make it possible for 3D reconstruction of these structures. Here we use SCC{\_}MEASURE in SSW package to analyze the 3D geometric properties of the structures. The results are shown in \fig{fig:3drec}, where the reconstructed points of P1, P2 and L1 are marked by blue cross symbols, cyan cross symbols, and red pluses, respectively. The geometric parameters of these relevant structures, obtained from the 3D reconstructions, are tabulated in \tbl{tbl:gp}: $D$ is the mean distance from the flare site, with the superscripts and subscripts indicating the spatial ranges of the structures, $H$ is the maximum 3D height, and $L$ is the estimated 3D length of the structures, which is calculated by summing the 3D distances of neighbouring reconstructed points by SCC{\_}MEASURE. It should be noted that $L$ only represents the length of the parts of the structures that could be clearly recognized in the FOVs of both SDO and STEREO. Obviously, the estimated $L$ is the lower limit of the actual length. The white curves in the left panels denote the solar limb as seen by SDO. It is obvious that P1, P2 and L2 were located near or to the west of (in front of) the solar limb in SDO, indicating that most parts of P1, P2 and L2 could be tracked in the 3D reconstruction. The coronal loops L1, however, were located more distant from the solar limb in SDO. As a result, some lower part of L1 were occulted by the solar disk in the perspective of SDO, which explains that its visible part in SDO has a minimum 3D height of 88 Mm. For comparison, the minimum tracked 3D heights of P1, P2, and L2 are only 15 Mm, 20 Mm and 6 Mm, respectively. Therefore, the calculated length from the 3D reconstruction for L1 is far from accurate; although the calculated lengths for P1, P2 and L2 are also underestimated, the deviations should not be large.

\subsection{Overview of the event}
\label{sec:overview}

\begin{figure}
\centerline{\includegraphics[width=1.0\textwidth]{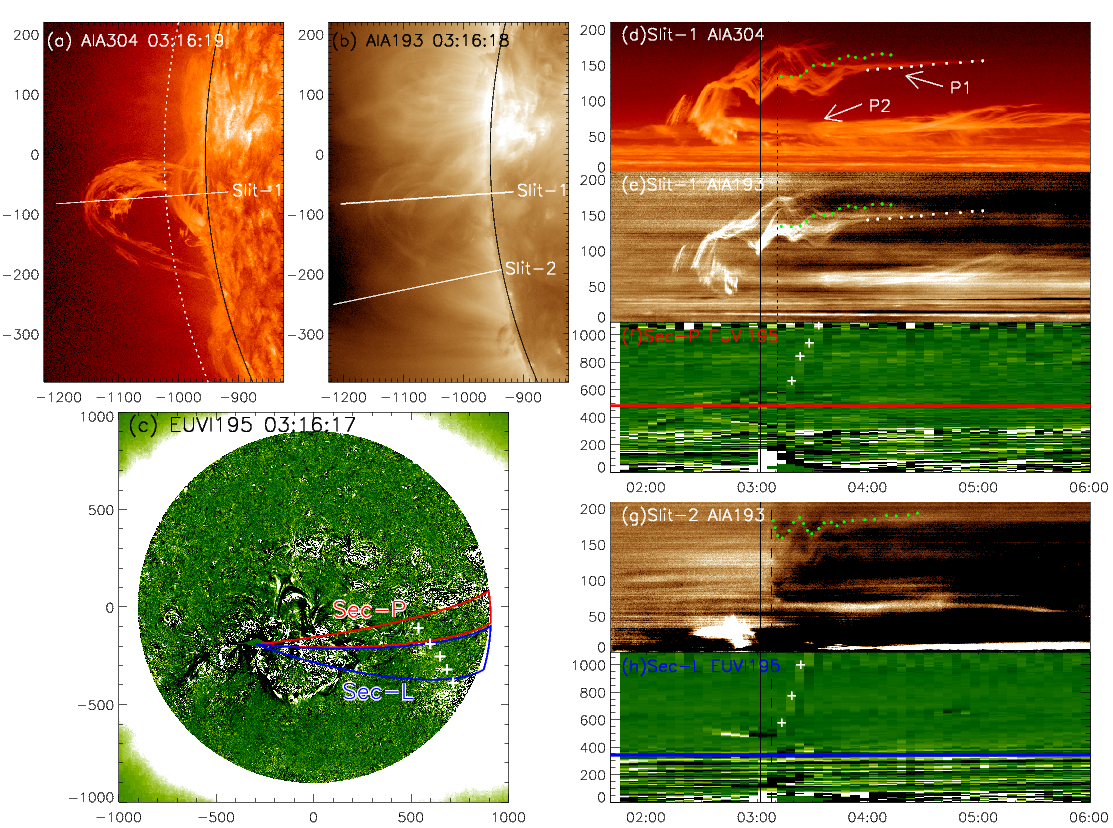}}
\caption{Analysis of the oscillations and the wave. Panels (a) and (b) show a 304~{\AA} image and an 193~{\AA} image, with two virtual slits Slit-1 and Slit-2 perpendicular to the solar surface; panel (c) is an 195~{\AA} running difference image, with two sectors region marked as Sec-P and Sec-L, whose centers are both located at the flare region, and the wave front is marked by the white cross symbols; panel (d) is the space$-$time stack plot obtained from original 304~{\AA} images along Slit-1 in panels (a), and panel (e) is from base difference 193~{\AA} images along slit 1; panel (g) is the stack plot from base difference 193~{\AA} images along Slit-2 in panel (b); panels (f) and (h) are stack plots from running difference 195~{\AA} images along Sec-P and Sec-L in panel (c), respectively. The wave structures are marked by cross symbols in panels (f) and (h). The green dots in panel (d), (e), and (g) represent the oscillating prominence thread and coronal loops. The red solid line in panel (f) and the blue one in panel (h) correspond to the locations of the prominence P1 and the coronal loops L1, respectively. The white dotted curve in panel (a) denotes the height of 50 Mm.}
\label{fig:slice}
\end{figure}

A big flare erupted at about 03:00 UT on 2012 September 22, corresponding to the vertical solid lines in \fig{fig:slice}(d)-\ref{fig:slice}(h). The flare generated a large-scale coronal wave, triggering the oscillations of the prominence P1 and the coronal loops L1 (also see the animation accompanying \fig{fig:slice}). The wave front is tracked in EUVI 195~{\AA} running difference image, as marked by the white cross symbols in \fig{fig:slice}(e). \fig{fig:slice}(f) and \ref{fig:slice}(h) are the space-time stack plots generated from EUVI 195~{\AA} running difference images in the sector regions Sec-P and Sec-L in \fig{fig:slice}(c), respectively. The centers of these sector regions are both located at the flare site. In comparison with \fig{fig:location}, it is revealed that Sec-P is across the region of the prominences, Sec-L the region of L1. The horizontal red and blue lines in \fig{fig:slice}(f) and \ref{fig:slice}(h) indicate the locations of the prominence P1 and the loops L1. The wave front, which is marked by the white cross symbols in \fig{fig:slice}(c), can also be recognized in \fig{fig:slice}(f) and \ref{fig:slice}(h), as marked by the white cross symbols. The propagating velocities of the coronal wave are obtained by linear fitting: (440$\pm$9) km s$^{-1}$ along Sec-P and (560$\pm$2) km s$^{-1}$ along Sec-L, indicating that it should be a fast wave. The velocity of the wave was almost constant during its propagation, so that the arrival times of the wave at the locations of P1 and L1 are calculated to be 03:11 UT for P1 and 03:07 UT for L1, denoted by the vertical dotted line in \fig{fig:slice}(d)-\ref{fig:slice}(f) and the vertical dashed line in \fig{fig:slice}(g)-\ref{fig:slice}(h), respectively. Both the prominence threads and the coronal loops began to oscillate with the wave passage, indicating that the oscillations were indeed triggered by the coronal wave.
\par

\subsection{Oscillating parameters}
\label{sec:parameters}

\begin{figure}
\centerline{\includegraphics[width=0.8\textwidth,clip=]{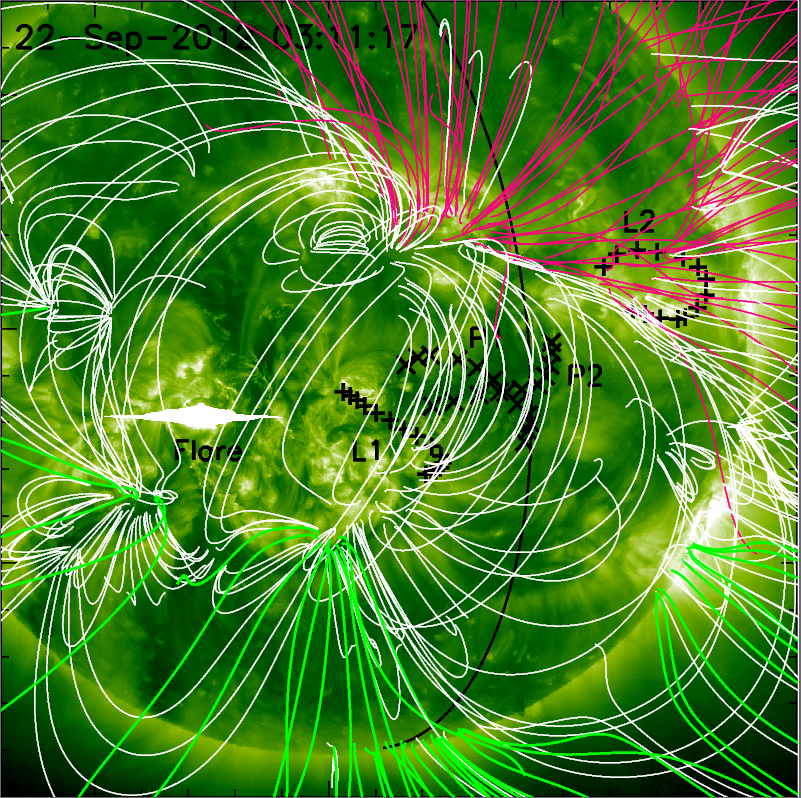}}
\caption{Magnetic field lines generated by PFSS extrapolations \citep{Schrijver2003a}. The white lines correspond to closed magnetic field lines; the red and green lines correspond to open magnetic field lines. The solid black curve denote the solar limb as seen by SDO. The prominences (P1, P2) are marked by the black cross symbols, and the coronal loops (L1, L2) by the black pluses.}
\label{fig:pfss}
\end{figure}

\fig{fig:slice}(d) and \ref{fig:slice}(e) are the space-time stack plots illustrating the motion in P1, generated from Slit-1 (see \fig{fig:slice}(a) or \fig{fig:slice}(b)). \fig{fig:slice}(d) is obtained from AIA 304~{\AA} original images, and \fig{fig:slice}(e) from AIA 193~{\AA} base difference images. During the pre-flare phase, P1 rose upwards almost as a whole, and then was suspended in the corona with an arch-like shape. The prominence was stabilized presumably by the overlaying arcades (see \fig{fig:pfss}), and prominence material was observed to fall back to the surface. Apart from the motion of the prominence as a whole, some threads of P1 also began to oscillate transverse to the prominence spine at about 03:11 UT, immediately after the wave passage. The oscillation of the prominence threads, marked by green dots, can be well-fitted with a damped cosine function with an initial height $h_0$, velocity $v_0$ and acceleration $a_0$:
\begin{equation}
h(t)=h_0+v_0t+\frac{a_0}{2}t^2+A\cos(\frac{2\pi}{T}t+\phi)\mathrm{e}^{-t/\tau},\label{equ:p1}
\end{equation}
where $A, T, \tau $ corresponds to the amplitude, period, and e-folding damping time, respectively. The height $h$ is measured from \fig{fig:slice}(d), and the error of $h$ is selected to be the length of 5 pixels in the image: $\bigtriangleup h\approx2.2$ Mm. The fitting yields that $v_0=(21\pm1)$ km~s$^{-1}$, $a_0=(6.2\pm0.6)$ m~s$^{-2}$, $A=(5.7\pm1.0)$ Mm, $T=(1000\pm20)$ s and $\tau=(2500\pm900)$ s. The velocity amplitude is calculated to be ($35\pm7$) km~s$^{-1}$, belonging to the large-amplitude oscillation. Apart from the damped oscillation, the non-oscillating part of this prominence was still undergoing slow rise, as marked by the white dots. With a linear fitting, the slow rise velocity is calculated as $v_0=(3.6\pm0.1)$ km~s$^{-1}$. Different from P1, P2 was non-oscillating, with no obvious motions along the slit, as shown in \fig{fig:slice}(d) and (e).
\par

\fig{fig:slice}(g) is the stack plot generated from AIA 193~{\AA} base difference images along Slit-2 in \fig{fig:slice}(b), which reveals the transverse oscillation of L1. The coronal loops also began to oscillate immediately after the the wave passage at about 03:07 UT. The coronal loops oscillated transversely as a whole, indicating a cylindrical kink mode (also see the animation accompanying \fig{fig:slice}). The oscillation of the coronal loop is outlined by the green dots in \fig{fig:slice}(g). A damped cosine function with a uniform slow rise velocity $v_0$ is used here to fit the oscillation of L1:
\begin{equation}
h(t)=h_0+v_0t+A\cos(\frac{2\pi}{T}t+\phi)\mathrm{e}^{-t/\tau}.\label{equ:l1}
\end{equation}
The fitting yields that $v_0=(4.8\pm0.2)$ km~s$^{-1}$, $A=(20\pm1)$ Mm, $T=(960\pm10)$ s and $\tau=(1300\pm100)$ s. The velocity amplitude is about ($130\pm10$) km~s$^{-1}$. The oscillating periods of the prominence threads and the coronal loops are approximately the same.
\par

\subsection{Estimation of the magnetic field}
\label{sec:esti-mag}

\begin{figure}
\centerline{\includegraphics[width=1.0\textwidth,clip=]{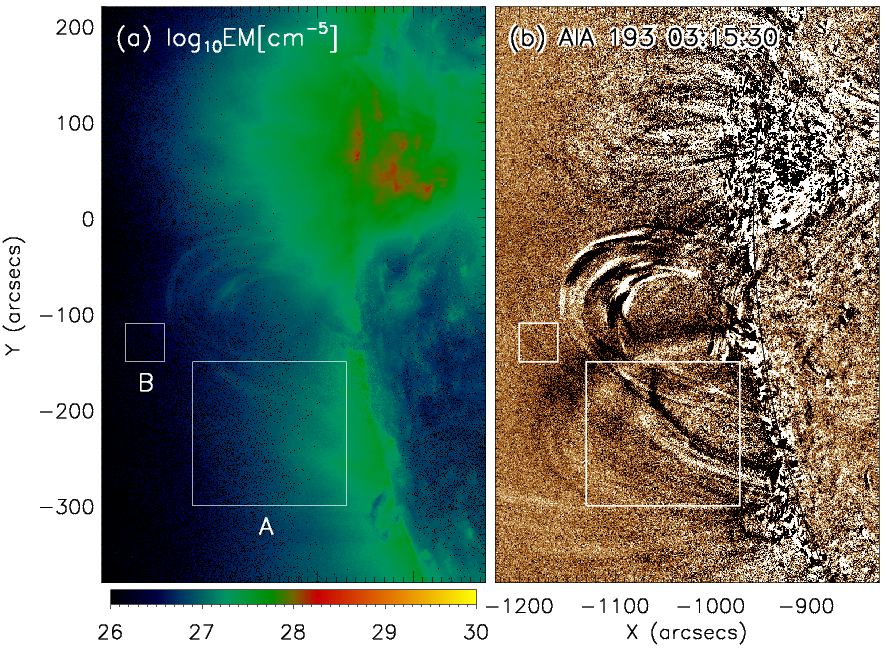}}
\caption{DEM results to calculate the densities. Panel (a) is the distribution of logarithmic EM; panel (b) is the running difference AIA 193~{\AA} image at the same time. The larger square marked as A in panel (a) and (b) denotes the region utilized to calculate the density around the oscillating coronal loops L1, and the smaller square marked as B is the calculating region for background corona. }
\label{fig:dem}
\end{figure}

In order to calculate the magnetic field from the oscillating parameters, the densities within the oscillating prominence and coronal loops should be estimated first, which is achieved by differential emission measures \citep[DEM;][]{Hannah2012a}. \fig{fig:dem} demonstrates the distribution of the emission measure (hereafter, EM), obtained from the integral of the DEM results. The relationship between EM and the electron density is written as:
\begin{equation}\label{equ:cal}
EM=n^2d,
\end{equation}
where $n$ is the electron density, and $d$ is the column depth along the line of sight \citep{Aschwanden2001a}. The distribution of logarithmic EM is shown in \fig{fig:dem}(a); the corresponding AIA 193~{\AA} running difference image is shown in \fig{fig:dem}(b). We select a square region around L1 marked as A in \fig{fig:dem}; the average EM within this region, $(8.1\pm0.9)\times10^{26}$~cm$^{-5}$, is used in Equation \ref{equ:cal} to calculate the density in the oscillating coronal loops L1. The error of EM is also calculated by DEM method. We assume that the LOS depth of the loop is the same as its width, which is estimated to be about 10 Mm by SCC$\_$MEASURE. Therefore, the electron density in L1 is calculated as: $n_L=(9.0\pm0.5)\times10^8$~cm$^{-3}$. Under the assumption that the corona is fully ionized, the mass density is $\rho_l=m_p n_L=(1.4\pm0.1)\times10^{-15}$~g~cm$^{-3}$, where $m_p=1.6\times10^{-24}$ g is the mass of a proton. 
\par
The method of calculating density through DEM can not be directly used for prominences. This is because main parts of prominences are observed only in low temperature 304~{\AA} waveband, which is optically thick, and not adequately treated by the CHIANTI model \citep{Woods2009a}, on which the DEM algorithm is based. The bright structure at the prominence site in \fig{fig:dem}(a) represents the high temperature corona material within the magnetic system of the prominence, i.e. only part of the prominence. \cite{Labrosse2010a} concluded that the prominence plasma is typically 100 times cooler and denser than its coronal surroundings. Based on this conclusion, we select another region, marked as B in \fig{fig:dem}, to calculate the density of the corona around the oscillating prominence, then the density in P1 can be estimated. The average EM within region B is $(1.3\pm0.1)\times10^{26}$~cm$^{-5}$. The column depth of the corona should be larger than that of the coronal loops. Here we use the pressure scale height of 1 MK plasma $H_p\approx$ 60 Mm as the column depth for the background corona. Then the electron density in the corona around P1 is calculated to be $n_C=(1.4\pm0.1)\times10^8$~cm$^{-3}$, and mass density $\rho_C=(2.3\pm0.2)\times10^{-16}$~g~cm$^{-3}$. Therefore, the mass density of the prominence is estimated to be $\rho_P\approx(2.3\pm0.2)\times10^{-14}$~g~cm$^{-3}$.

From the densities obtained above, the local magnetic field within the oscillating structures can be estimated. \cite{Kleczek1969a} proposed a model describing the oscillations of a prominence, where the prominence is considered as a bundle of magnetic plasma threads anchored in the photosphere, and the restoring force is considered as magnetic tension. The oscillating period of the prominence P1 is given by
\begin{equation}
T=2\pi LB^{-1}\sqrt{\pi\rho},
\end{equation}
where $L$ is the length of the oscillating prominence threads, $\rho$ the mass density, and $B$ the strength of the effective magnetic field providing the restoring force. From the measured value  $T=(1000\pm20)$ s and $L=220$ Mm, B is calculated as ($37\pm2$) Gauss, with the estimated density $\rho_P\approx(2.3\pm0.2)\times10^{-14}$~g~cm$^{-3}$. The corresponding Alfv\'{e}n speed is ($690\pm70$) km~s$^{-1}$, calculated by $V_A = B/\sqrt{4\pi\rho}$. Note that since the value of $L$ in \tbl{tbl:gp} is an underestimation of the length (see \sect{sec:3d}), the calculated B should be the lower limit of the strength of the magnetic field.
\par
By using the phase speed of the fast kink mode, \cite{Nakariakov2001a} calculated the local magnetic field with the parameters of the oscillating coronal loops as:
\begin{equation}
B=\left(4\pi\rho_0\right)^{1/2}V_A=\frac{\sqrt{2}\pi^{3/2}L}{T}\sqrt{\rho_0(1+\rho_e/\rho_0)},\label{equ:kink_speed}
\end{equation}
where $\rho_0$ and $\rho_e$ are the internal and external densities of the coronal loops, $V_A$ is the Alfv\'{e}n speed, $L$ is the length of the loop and $T$ is the period. As discussed in \sect{sec:3d}, the length of L1 can not be directly measured from the 3D reconstructions. The estimated length of the 3D reconstructed part of L1 is $L_{\mathrm{L1}}=220$ Mm (see \tbl{tbl:gp}), and the heights of the two `footpoints' of the reconstructed part are 88 Mm and 122 Mm, indicating the length of L1 should be at least 430 Mm. Here we use $L=$ 430 Mm, the lower limit, for L1. Assuming the external to internal density ratio to be 0.1 \citep{Nakariakov2001a}, the magnetic field and corresponding Alfv\'{e}n speed are calculated as ($14\pm1$) Gauss and ($1000\pm110$) km~s$^{-1}$, with $T=(960\pm10)$ s and $\rho_l=(1.4\pm0.1)\times10^{-15}$~g~cm$^{-3}$.
\par
As mentioned above, since the coronal loops L1 oscillated in a cylindrical kink mode, the damping time scale of the oscillation can also be used to calculate the local physical parameters. By comparing the the damping time scaling predicted by several damping mechanisms with that of the transverse oscillations in 26 coronal loops in 17 events, \cite{Ofman2002a} demonstrated that the damping power index predicted by phase mixing is in excellent agreement with the observation, superior to other mechanisms. Then according to the phase mixing model deduced by \cite{Roberts2000a}, the e-folding damping time of the oscillation $\tau_{\mathrm{decay}}$ is given by
\begin{equation}
\tau_{\mathrm{decay}}=\left(\frac{6L^2l^2}{\nu\pi^2V_A^2} \right)^\frac{1}{3},\label{equ:phase_mixing}
\end{equation}
where $L$, $l$, $\nu$, $V_A$ correspond to the length of the coronal loop (the same as that in Equation \ref{equ:kink_speed}), scale of the inhomogeneity across the loop, the coronal viscosity, and Alfv\'{e}n speed, respectively, and $\tau_{decay}$ for L1 has been calculated to be ($1300\pm100$) s (see \sect{sec:parameters}). By assuming $l=0.01L$ and $\nu=4\times10^{13}~$cm$^2~$s$^{-1}$ \citep{Roberts2000a}, the Alfv\'{e}n speed $V_A$ is calculated to be ($480\pm60$) km~s$^{-1}$, and the corresponding magnetic field is ($6.4\pm0.9$) Gauss, comparable to the field strength given by the phase speed of the fast kink mode. The difference between the estimations of the magnetic fields by the phase speed of the fast kink mode and by phase mixing might result from the underestimation of the length of L1 ($B\sim L$ in Equation \ref{equ:kink_speed} and $B\sim V_A\sim L^2$ in Equation \ref{equ:phase_mixing}, so that with increasing $L$, the difference between those calculated from Equation \ref{equ:kink_speed} and Equation \ref{equ:phase_mixing} will decrease).

\subsection{Propagating height of the wave}
\label{sec:height}
Since the oscillations are triggered by the coronal wave, the different responses to the wave passage of different structures in different locations sheds light on the nature of the coronal wave. Besides oscillating P1 and L1, there also existed two non-oscillating structures P2 and L2 (see \sect{sec:3d} and the animation accompanying \fig{fig:slice}). In order to analyze the propagation of the wave, potential-field source-surface \citep[PFSS;][]{Schatten1969a} extrapolation is introduced to illustrate the magnetic configurations around the structures of interest. Full-sun PFSS extrapolation results at 04:00 UT on 2012 September 22 are shown in \fig{fig:pfss}. Since the relevant structures were located near the solar limb, PFSS extrapolations at the regions of these structures are based on data a week later or two weeks earlier. On the other hand, the magnetic fields around the prominences and the loops might be non-potential. As a result, PFSS results deviate from the detailed observed 3D structures. Here, we only use PFSS to trace magnetic field lines at high altitude, which represents the external large-scale magnetic field over the relevant structures. Based on the traced external magnetic configuration, we could analyze the propagation of the wave. \cite{Wang2000a} concluded that coronal waves are deflected away from active regions and coronal holes, where the velocity of fast-mode magnetohydrodynamic wave is large. Thus, the PFSS result reveals that the coronal wave could propagate through the region of L1, P1 and P2. However, L2 was located at the region surrounded by open magnetic fields, which will prevent the coronal wave passage, resulting in the non-oscillating behavior of L2.
\par
\fig{fig:pfss} reveals that P2, although at a lower altitude (see \tbl{tbl:gp}), was located at almost the same region as P1, under the same group of magnetic arcades. This indicates that the responses of P1 and P2 to the wave should not vary that much. As shown in \fig{fig:slice}(d), however, the threads in P1 began to oscillate after the wave passage, whereas there were no obvious motions along slit-1 in P2. As demonstrated above, large-amplitude oscillations in P1 was triggered by the wave. This indicates that the compression of the wave should be strong. Thus, if the wave had propagated through P2, at least disturbances along slit-1, i.e. the same direction as the oscillation in P1, should be triggered in P2 by the wave. Note that both P1 and P2 were located near the solar limb from the perspective of SDO, so that the projection effect is small. The fact that no disturbances were detected along the slit in P2 after the wave passage indicates that there might be no interaction between the wave and P2. This should result from the different altitudes of P1 and P2. Therefore, we may conclude that the wave must propagate above a certain height, i.e. the lower boundary of the wave front is between 50 Mm and 130 Mm, the maximum heights of P2 and P1, respectively. Similar conclusion was also reached by \cite{Liu2013a}, in which the lowest of the four filaments did not respond to the Moreton wave. By triangulations of the wave front, \cite{Patsourakos2009a} found that the height of the wave above the solar surface is about 90 Mm. \cite{Kienreich2009a} also suggested that the coronal wave originates from 80-100 Mm above the solar surface with the STEREO quadrature observations. Different from those studies, the flare site in this paper is far behind the solar limb from the Earth perspective. As a result, the wave front can hardly be observed in AIA images, so that the methods utilized above are infeasible here. Our result, obtained from the interactions between the wave and the prominences, is consistent with previous studies. Such heights are comparable to the coronal scale heights for quiet Sun, which is 50$\sim$100 Mm for the temperatures of 1$\sim$2 MK. \cite{Patsourakos2009a} concluded that the fast-mode wave perturbs the ambient coronal plasma with its bulk confined within a coronal scale height, also indicating that our observation is consistent with the fast-mode wave propagation.
\par

\section{Discussion and conclusion}
\label{sec:conclusion}
As discussed above, the oscillations in P1 and L1 were triggered by the coronal wave. The exact nature of the relationship between the properties of the wave and the filament activation is currently not well understood \citep{Tripathi2009a}. However, we may still conclude that the wave energy deposited in P1 and L1 should be no less than the oscillating energies of P1 and L1, respectively. The oscillating energies can be estimated from the oscillating parameters:
\begin{equation}
E=\frac{1}{2}mv_{max}^2=\frac{2\pi^2mA^2}{T^2}, ~m=\rho\pi r^2L
\end{equation}
where $m$, $A$, $T$, $\rho$, $r$, $L$ correspond to the mass, amplitude, oscillating period, mass density, radius, length of the prominence (coronal loops), respectively. Assuming $r=10$ Mm for the prominence threads, the same as that of L1, the oscillating energies can be estimated from the measured values of $A$, $T$, $L$ and $\rho$ in \sect{sec:parameters} and \sect{sec:esti-mag}: the oscillating energy of the prominence P1 is $E_{\mathrm{P1}}=(1.0\pm0.4)\times10^{28}$~erg and the energy of the loops L1 is $E_{\mathrm{L1}}=(1.6\pm0.3)\times10^{28}$~erg. Therefore, the lower limit of the dissipated wave energy within the region of P1 and L1 is $\bigtriangleup E=E_{\mathrm{P1}}+E_{\mathrm{L1}}\sim10^{28}$~erg. P1 and L1 span about 30 deg with respect to the flare, as is shown by the black sector in the bottom left panel in \fig{fig:location}, and only the wave within this sector interacted with P1 and L1. Since coronal waves are generally considered to be isotropic and propagate in a wide-range sector almost symmetrically relative to the flare site \citep{Chertok2003a,Warmuth2001a,Nakariakov2005a}, the total energy of the wave should be much larger than the deposited wave energy within P1 and L1, probably of the order of $10^{29}\sim10^{30}$~erg.
\par

\begin{figure}
\centerline{\includegraphics[width=1.0\textwidth,clip=]{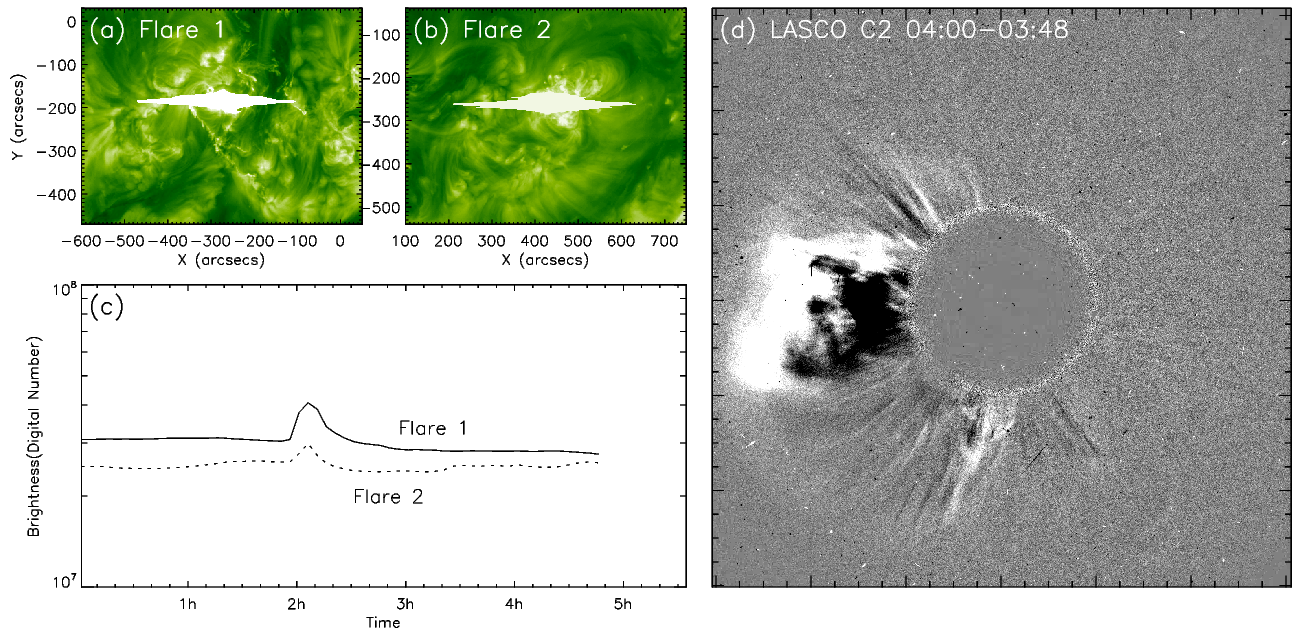}}
\caption{The lightcurves of the flares. Panel (a) is the image of the flare discussed in this paper (Flare 1), whose lightcurve corresponds to the solid curve in  panel (c). Panel (b) shows a M9.0 class flare as comparison (Flare 2), whose lightcurve is the dotted curve in panel (c). Panel (d) is the running difference image of LASCO C2 observations for the CME generated by the Flare 1.}
\label{fig:lightcurve}
\end{figure}

Since the relevant flare is behind the solar limb, there is no GOES observations for it. \fig{fig:lightcurve}(c) exhibits the comparison between the lightcurves of the flare of interest (Flare 1) and an M9.0 flare on 2012 October 20 (Flare 2), in order to roughly estimate the class of the relevant flare. These lightcurves are both generated from EUVI 195~{\AA} direct images within a 650$\times$500 arcsecs region, as shown in \fig{fig:lightcurve}(a) and \ref{fig:lightcurve}(b). Although the backgrounds of these two flares were different, the increase of the peak value during Flare 1 relative to the background (the average value before the onset of the flare) is $1.0\times10^7$ dn, and that during Flare 2 is $3.7\times10^6$ dn. It is obvious that the M9.0 flare should be less energetic than Flare 1, indicating that Flare 1 might be a large flare, probably an X class one. Thus, the total energy released by the relevant flare should be of the order of $E_{flare}=10^{32}$~erg \citep{2009psfbook}. \cite{Aschwanden2014a} also demonstrated that the magnetic free energies of large flares are usually larger than $10^{31}$~erg. Therefore, the flare energy is much larger than the estimated wave energy. The direct trigger for coronal waves, at least those have a bright wave front, are usually considered to be Coronal Mass Ejections (CMEs) \citep{Biesecker2002a,Ballai2005a,Chen2006a}. In this event, a CME was generated by the flare, as shown in \fig{fig:lightcurve} (d), the observation from the Large Angle and Spectrometric Coronagraph \cite[LASCO;][]{Brueckner1995a} on board the Solar and Heliospheric Observatory (SOHO). From the the LASCO CME list on CDAW site, the kinetic parameters of the CME is as follows: the velocity of the CME is 774 km~s$^{-1}$ and the kinetic energy is $2.7\times10^{31}$ erg, also much larger than the estimated wave energy. Therefore, we conclude that the flare and the corresponding CME should be energetic enough to trigger the coronal wave.
\par
In summary, we investigate the oscillations, triggered by a global coronal wave, of the prominence and the coronal loops quantitatively. From the observed oscillating properties, local physical parameters are obtained. The magnetic field strength and the Alfv\'{e}n speed of the prominence are at least about 37 Gauss and 690 km~s$^{-1}$, those of the coronal loops 6$\sim$14 Gauss and 480$\sim$1000 km~s$^{-1}$. By comparing the locations and heights of the oscillating and non-oscillating structures, the propagating height of the wave is estimated to be 50$\sim$130 Mm, comparable to the coronal scale heights for quiet Sun. Finally, the lower limit of the energy dissipations of the coronal wave are roughly gauged by the oscillating energies, and the relevant flare and CME are proved to be energetic enough to trigger this coronal wave. The 3D reconstructions play an important role in analysing the observations. Oscillations can be used in not only coronal seismology, but also revealing the properties of the wave.
\par
This research is supported by Grants from NSFC 41131065, 41574165, 41421063, and 41304134, MOEC 20113402110001, CAS Key Research Program KZZD-EW-01-4, the fundamental research funds for the central universities WK2080000077, and the foundation for Young Talents in College of Anhui Province (2013SQRL044ZD). The CME catalog used to obtain the kinetic parameters of the relevant CME is generated and maintained at the CDAW Data Center by NASA and The Catholic University of America in cooperation with the Naval Research Laboratory. SOHO is a project of international cooperation between ESA and NASA.

\end{document}